\documentclass[11pt]{article}
\usepackage[utf8]{inputenc}
\usepackage{amsmath, amssymb, amsthm}
\usepackage{geometry}
\usepackage{hyperref}

\usepackage{color}

\usepackage[most]{tcolorbox}

\usepackage{graphicx,url}
\usepackage{subfigure}

\title{\textbf{Propositional Measure Logic}}
\author{Francisco Aragão \\ \small{Federal University of Ceará} \\ \small{\texttt{aragao@ufc.br}}}
\date{}

\usepackage{amsthm}  

\newtheorem{theorem}{Theorem}
\newtheorem{definitionMinha}[theorem]{Definition}
\newtheorem{definition}[theorem]{Definition}
\newtheorem{aragao}[theorem]{Example}

\definecolor{antiquebrass}{rgb}{0.8, 0.58, 0.46}
\definecolor{auburn}{rgb}{0.43, 0.21, 0.1}
\definecolor{alizarin}{rgb}{0.82, 0.1, 0.26}
\definecolor{alizarinNew}{rgb}{0.72, 0.1, 0.16}
\definecolor{cinzaclaro}{rgb}{0.78, 0.78, 0.78}
\definecolor{cambridgeblue}{rgb}{0.64, 0.76, 0.68}

\begin{document}

\maketitle

\begin{abstract}
We present a propositional logic with fundamental probabilistic semantics, in which each formula is given a real measure in the interval $[0,1]$ that represents its degree of truth. This semantics replaces the binarity of classical logic, while preserving its deductive structure. We demonstrate the soundness theorem, establishing that the proposed system is sound and suitable for reasoning under uncertainty. We discuss potential applications and avenues for future extensions of the theory. We apply probabilistic logic to a still refractory problem 
in Bayesian Networks.

\end{abstract}

\section{Introduction}

Classical formal logic underpins mathematics, computer science and formal inference. However, to deal with ambiguity and partial information, new approaches have emerged - examples of which are fuzzy logic, probabilistic modal logic, Bayesian networks and belief-based systems.

Even though progress has been made, these approaches generally have a limitation: 
the probability or degree of belief, in general, being kept out of the logical semantics, remaining at another level of interpretation on a deterministic model. 
In other words, maintaining the binary characteristic of truth - true or false, with uncertainty being treated as associated with models, rather than a property of logical language in itself. The proposed logic will be used to solve the problem of tackling certain types of uncertainty and imprecision with Bayesian Networks. The aim is to take advantage of the conceptual and practical benefits of this system in practical situations that have not yet been adequately explored.

We introduce an extension to classical propositional logic which follows Hilbert's axiomatic model, which is correct and quasi-complete. The syntactic aspect of classical propositional language remains intact, though the semantics is given by \textit{beliefs}, e.g. a degree of truth, to which each formula is assigned, that is, a measure in the real interval [0, 1]. This approach shows ability to deal with applications from theory of knowledge to artificial intelligence. We develop a unified and principled logic for reasoning under uncertainty, including formalism, as well as expressiveness and interpretability. We have proven that the proposed probabilistic semantics is sound in relation to classical logic: every classically valid formula has positive probability.

\section{\textcolor{black}{The Propositional Measure Logic - PML}}\label{sec:PMLandTheSolution}

\textcolor{black}{
In this section, we introduce a probabilistic logic that aims to work on logical reasoning and probabilistic reasoning, called PML.
It is defined semantically in the Definition \ref{def:justificacaoBis}, based on a generalization of the concept of truth valuation, the well-formed sequence of truth valuations,
Definition \ref{def:SequenceBemFormada}.
}

The union of Logic and Probability can be traced back to the 17th century. A comprehensive list of pioneering approaches can be found in 
\cite{hailperin1984}, \cite{pearl1988} and \cite{hailperin1991}.
We introduce the Probabilistic Epistemic Logic 
whose language differentiates between `safe' knowledge and `inaccurate' knowledge. The latter may be an affirmative testimony, while the former can be a truth accepted on the basis of solid, consistent, argumentation.

Probabilistic Epistemic Logic differs from traditional logic, such as Nilsson's [Nil86] and Hailperin's [Hail84] probabilistic logic because it has extended the truth function concept. As an ultimate motivation, we aim an appropriated probabilistic logic to fusion imprecise belief.

\textcolor{black}{
The intuition behind Probabilistic Epistemic Logic 
is to measure the Probability of a sentence, you need a semantic concept whose structure is as sophisticated as the sentence structure itself. 
With a probabilistic semantics, it will be possible to make decisions in situations where purely logical systems, classical or non-classical, do not.
}

We can handle contradiction without trivialization by doing logical reasoning, and also we
can make decisions in a probabilistic-like way. The proposed system accommodates reasons and measurements for imprecise information.
We proceed by introducing the basic notion of a formula:

\begin {definition} \label{def:Formula} [Formula]\cite{VanDalen2004}: 
Given a set of propositional letters, we define recursively a Well Formed Formula (\textit{WFF}) as
\textcolor{black}{
\begin{itemize}
    \item $\textrm{Propositional letters are formulae};$
    \item $\textrm{If } \alpha \ \textrm{ and } \ 
    \beta \textrm{ are formulae, then }   
    (\alpha \vee \beta) \ and \ (\alpha \rightarrow \beta) \textrm{ are formulae}$
    \item $\textrm{If } \alpha \textrm{ is a formula, then } (\neg \alpha) \textrm{ is also a formula}$.
\end{itemize}
}

\end {definition}

A \textit{truth valuation} is a map from the set of all formulae from a logical language into the set $\{true, false\}$.
In order to elaborate a probabilistic-like semantics to the logical language presented in Definition \ref{def:Formula}, we need a structured extension of truth valuations that proves to be isomorphic to the structure of those formulas.
Section \ref{sec:WellFormedSequencesOfTruthFunctions} formalizes the \textit{Well Formed Sequences of truth valuations}:

\subsection{Well Formed Sequences of truth valuations}\label{sec:WellFormedSequencesOfTruthFunctions}

\textcolor{black}{
The set of all possible truth valuations determining the truth of a sentence does not have the necessary structure to define a measure over a language. We are obliged to define a new truth object called \textit{well-formed sequence}. Traditionally, logical truth is established through a meta-linguistic relationship between a simple object (a function) and a complex object (a proposition).
We propose to replace this 
by the confrontation of two complex objects: a proposition, as before, but instead of the simple truth valuation, we would have a structured sequence of truth valuations.}

\textcolor{black}{To define a \textit{mathematical measure} over a language, it is necessary to capture some idea of quantity, as Nilsson and Hailperin among others have tried. See Section \ref{sec:related-work}.   
Each structured sequence represents a different way of being true for a logical formula, thus we define:}

\begin {definition} \label{def:SequenceBemFormada} 
[Well-Formed Sequence]: 
Let ${C}$ be a collection of truth valuations.
We say that $ s $ is a Well Formed Sequence (\textit{WFS}) generated from a collection ${C}$ of truth valuations when
\begin {itemize}

\item If $ v $ is a \textit {truth valuation} then $\big< v \big>$ is a well-formed sequence.

\item If $s_i$ and $s_j$ are well-formed sequences, then $\big< s_i , s_j \big>$ is also a well-formed sequence.

\item Nothing else is \textit{WFS}.

\end{itemize}
\end {definition}

We call \textbf{\textit{WFS}(${C}$)} the set of all well-formed sequences generated by 
the collection ${C}$ of truth valuations.
 Each language formula partitions the universe of well-formed sequences by two, those \textit{WFS} associated to the formula and those not associated:

\begin{definitionMinha}\label{def:associacao}
[Association]: Let $ \varphi $ be a \textit {\textit{WFF}} and $ s $ be a \textit {WFS}.  
We say that $ s $ is associated to $ \varphi $, (notated $\varphi_s$) when 

if $\varphi = A$ and $s = \big< v \big>$, then $\varphi$ is associated to $s$;

\textcolor {black}{
if $\beta$ is associated with $s$, then $(\neg \beta)$ is also associated with $s$,
}

\textcolor {black}{
if $\alpha$ is associated with $s_i$ and if $\beta$ is associated with $s_j$,
then $\varphi = (\alpha \vee \beta)$ is associated with $s = \big< s_i , s_j \big>$.
}

\textcolor {black}{
if $\alpha$ is associated with $s_i$ and if $\beta$ is associated with $s_j$,
then $\varphi = (\alpha \rightarrow \beta)$ is associated with $s = \big< s_i , s_j \big>$.
}

We call ${A}_\varphi$  the set of all \textit{WFS} associated to $\varphi$. 
\end{definitionMinha}

Sequences associated with an atom $ A $ are of the form $v$, where $v$ is a truth valuation.
Formulae constructed by binary connective as $(\alpha \vee \beta)$ are associated with sequences of type $\big< s_1 , s_2 \big>$, where $s_1$ and $s_2$ are well-formed sequences.
Negated formulae as $( \neg \alpha )$ are specially associated to the same type of sequences as those associated to $\alpha$.
Next, we develop a probability measure that will work as a logical semantics. 

\subsection{Probabilistic semantics - Mathematical Measures as Logical Semantics}\label{sec:ProbabilisticSemantics}

We introduce the \textit{justification} between a sequence and an associated formula.
Some background is necessary.
(i) The collection ${C}$ of truth valuations provides the elements with which well-formed sequences are built. Let $ S ^ * $ be the infinite enumerable set of all \textit{WFS} generated from ${C}$.
(ii) Each formula $ \alpha $ partitions $ S ^ * $ into associated and non-associated \textit{WFS} (see Definition \ref{def:associacao}). The set ${A}_\alpha$ of $\alpha$-associated \textit{WFS} is finite and the set of non-associated \textit{WFS} is infinite enumerable.

Truth is determined in classical logic by means of truth valuations.
In Probabilistic Epistemic Logic (PML), the truth is determined by inspecting sequences of truth valuations.
This enables us to identify the subsets of truth valuations able to assure the truth of the sentence, regardless of the rest of the collection of \textit{WFS}s. 
When a subset of truth valuations in a structured form of a \textit{WFS} is capable of ensuring the veracity of the sentence, we say it justifies the
sentence. Definition \ref{def:justificacaoBis} formalize that idea:

To provide a foundation for the connection between a sequence and an associated formula, some contextual information becomes essential. (i) The assembly of truth valuations, denoted as collection ${C}$, serves as the fundamental building blocks for well-formed sequences. Let $S^*$ represent the infinite and enumerable set comprising all well-formed sequences generated from the collection ${C}$. 
(ii) Each formula, designated as $\alpha$, divides the set $S^*$ into two categories: associated well-formed sequences and non-associated well-formed sequences (Definition 4.3). The grouping of $\alpha$-associated well-formed sequences, termed $A_{\alpha}$, is finite, while the set of non-associated well-formed sequences remains infinite and enumerable.

In classical logic, the determination of truth is contingent upon truth valuations. However, in the context of PML, truth is established by examining sequences of truth valuations. This approach empowers us to identify subsets of truth valuations capable of ensuring the validity of a given statement, independent of the remaining collection of well-formed sequences. When a structured subset of truth valuations within a well-formed sequence can effectively guarantee the accuracy of the statement, we refer to it as the justification for that statement. 
This concept is formally build by Definition \ref{def:justificacaoBis}. 

\begin{definitionMinha}\label{def:justificacaoBis}
\textcolor{black}{
		[justification]: Let ${C}$ be a non-empty set of classical truth valuations, let $\alpha$ and $\beta$ 
		be formulae and 
		 $ s_i $ and $ s_j $
		 be well-formed sequences associated with $\alpha$ and $\beta$, respectively. 
         }
	
\noindent \textcolor{black}{ 
$|\!\!\!\approx\!\!(v   , A)$ iff
$|\!\!\!\approx\!\!( v   , A)$ iff
$v(A) = 1$}, \textcolor{black}{Where $A$ is a propositional letter.}

\noindent $|\!\!\!\approx\!\!(
\big< s \big>
, \neg \alpha)$ 		
		iff $|\!\!\! \not \approx\!\!(s , \alpha)$ 

\noindent $|\!\!\!\approx\!\!(\big< s_i , s_j  \big> , \alpha \rightarrow \beta)$
       iff 
       $|\!\!\! \not \approx\!\!(s_i  , \alpha)$ or $|\!\!\! \approx\!\!(s_j , \beta)$		
	
\noindent \textcolor{black}{$|\!\!\!\approx\!\!(
		\big< s_i , s_j \big> , 
		\alpha \vee \beta)$ iff $|\!\!\! \approx\!\!(s_i , \alpha)$ or $|\!\!\! \approx\!\!(s_i , \beta)$
}		
	
\noindent \textcolor{black}{ For any formula $\alpha$ in the language of PML
and any $\alpha$-associated sequence $s_\alpha$,
the justifiability of $\alpha$ by $s_\alpha$ is given by $|\!\!\!\approx\!\!(s_\alpha , \alpha)$.
}	
\end{definitionMinha}

\newcommand{\hlc}[2][yellow]{{
    \colorlet{foo}{#1}
    \sethlcolor{foo}\hl{#2}}
}

We denote the set of all sequences justifying a formula $\varphi$ as ${J}_\varphi$. 
The \textit{association} relationship between sequences (\textit {WFS}) and formulae (\textit {WFF}s) is \textbf {syntactic}, while the \textit {justification} relationship between them 
is \textbf {semantic}.
Example \ref{Ex:justificationExemploUnico} illustrates the concept of justification and association.

\begin{aragao}\label{Ex:justificationExemploUnico}(justification and association) Let $\varphi = (A \wedge \neg B)$ be a formula and let
\textcolor {black}{ ${C} = \{ v_0, v_1, v_2, v_3 \} $ be the standard collection of truth valuations  
for sentences with up to two \textit{atoms} such that 
$v_0(A) = v_0(B) = false$; 
$v_1(A) = false \textrm{ and } v_1(B) = true$;
$v_2(A) = true \textrm{ and } v_2(B) = false$ and
$v_3(A) = true \textrm{ and } v_3(B) = true$.
}
According to the Definition \ref{def:associacao} the sequences associated with $ (A \wedge \neg B) $ are of the form 
$s = \big<  v_i \ , v_j\big> $ with
$\big<  v_i \big> $ associated $A$ and $\big<  v_j\big> $ associated with $\neg B$.
Thus, by Definition \ref{def:justificacaoBis},
$\varphi$ is justified by 
\textcolor{black}{those 
$ 
s = \big<  v_i \ ,
v_j\big> $ where $v_i(A) = 1$ and $v_j(B) \neq 1$.
}
They are
\textcolor{black}{
$
\big< v_2 ,  v_1 \big>\ ,
\big<
 v_2 ,  v_3 
\big>\ , \ 
\big< v_3 ,  v_2 \big>\ ,
\big<
 v_3  , v_3  
\big>$\}.}

\end{aragao}

We define a measure maintaining our language in a normal form where all negation symbols are confined to be forming literals, appearing only before of a propositional letter.

\begin{definitionMinha}\label{def:PML}
	[PML]: Definimos PML como o par $(L \ , \ 
|\!\!\!\approx\!\! \
)$, em que
$L$ é a linguagem da Definição \ref{def:Formula} e $\ |\!\!\!\approx\!\! \ $ é a relação da Definição \ref{def:justificacaoBis}.

 \end{definitionMinha}

The PML is defined as pair A, where L is the language of Definition 1 and B is the relation of Definition 2.
Two technical definitions before presenting our mathematical measure over a logical langage:  

\begin{definitionMinha}\label{def:sigmaalgebra}
	[$\sigma$-algebra]\cite{bjork2009arbitrage}: Let $X$ be a set. A $\sigma$-algebra is a nonempty collection $\Sigma$ of subsets of $X$ closed under complement, countable union, and countable intersection. $X$ is called a sample space, and the pair $(X , \Sigma)$ is called a measurable space.

\end{definitionMinha}

\vspace{0.4cm}

\begin{definitionMinha}\label{def:measure}
[Measure]\cite{bjork2009arbitrage}: Let $\Sigma$ be a $\sigma$-algebra over a set $X$. A function $\mu$ from $\Sigma$ to the real line is a measure if 

Non-negativity: For all $E \in  \Sigma$, we have  $\mu (E)\geq 0$. 

Null empty set:  $\mu (\emptyset) = 0$. 

Countable additivity (or  $\sigma$-additivity): For all countable collections  $\{E_{k}\}_{k=1}^{\infty }$ of pairwise disjoint sets in $\Sigma, \ \ 
 \mu \left(\bigcup _{k=1}^{\infty }E_{k}\right)=\sum _{k=1}^{\infty }\mu (E_{k})$.

\end{definitionMinha}

We define a measure taking logical atoms as \textit {elementary aleatory events} and compound formulae as \textit{composite aleatory events}.
Random phenomena will be described in logical language, and no longer in the language of set theory. 
At \ref{def:LogicalMeasure}, we present a \textbf{logical measure} to deal with logical expressions .

\vspace{0.4cm}

\begin{definitionMinha}\label{def:AleatoryEvent}
	[Aleatory Elementary Event of a formula]: Let $\alpha$ be a formula and ${C}$
be a collection of truth valuations.  
    An Aleatory Elementary Event of $\alpha$ is a well-formed sequence generated
    from ${C}$ and \textit{associated} to $\alpha$.
\end{definitionMinha}

\vspace{0.4cm}

\begin{definitionMinha}\label{def:SampleSpace}
		[Sample Space for a formula]:  
Let $\varphi$ be a formula, and 
$\Omega = \{{s_{1}}_{\varphi}, ..., {s_{n}}_{\varphi}\}$ be the set of all well formed sequences associated to $\varphi$. We call $\Omega$ a \textit{sample space} for $\varphi$.
\end{definitionMinha}

\begin{aragao}\label{ex:ElementaryAleatoryEventOfaFormulaDISJUNCAO}(
Sample Space of a formula) Let $\varphi = (A \vee B)$. 

The sample space is the collection 
$ {A_{(A \vee B)}} = \{ \big<v_i , v_j\big> \ | \ v_i(A) = 1 \textrm{ or } v_j(B) = 0\}$.
According to the Definition \ref{def:associacao} they are 
$ {A}_{(A \vee B)} = \{
\big< v_2 \wedge  v_1 \big>\ ,
\big< v_2 \wedge  v_3 \big>\ , \ 
\big< v_3 \wedge  v_2 \big>\ ,
\big< v_3  \wedge v_3  \big>$
\}.

\end{aragao}

Our focus lies in events related to the truth of a formula within the sample space. These events take the form of well-formed sequences (\textit{WFS}) that substantiate, justify said formula. The ratio between the sequences that offer justification for the formula, compared to those associated with it, establishes the measure of the formula.

Truth can manifest in various interpretations. Each truth valuation signifies a perspective on a given sentence. Withal, 
the distribution of weight across truth valuations extends to encompass well-formed sequences of these functions. These sequences serve as the conduit for transferring measures from sequences to sentences. 
From
a basic weight distribution over the collection of 
truth valuations ${C}$, 
we establish a weight assignment for sets of well-formed sequences according to Definitions \ref{x} through \ref{def:LogicalMeasure}.
A formula's measurement is contingent on the well-formed sequences that validate it
and is done by means of \textit{well-formed sequences} that justify it.

\vspace{0.4cm}

\begin{definitionMinha}\label{x}
		[Weight Allotment]: A Basic weight distribution over the collection of truth valuations ${C}$ is a map $\Bar{p}$ in the interval $[0 , 1]$ such that $\Sigma_{ \{v \in {C}\} } \Bar{p}(v) = 1$.
Let $\Bar{p}(\cdot)$ be a BPD over ${C}$. 
We define the Weight Allotment $p$, as an extension of $\Bar{p}$ over the domain of all well-formed sequences as 
\begin{displaymath}
p(s) =
\left\{ 
  \begin{array}{ccc}
  \Bar{p}(v) & \textrm{if} & s =  v  \\
  p(s_i) \cdot p(s_j) & \textrm{if} & s = \big< s_i \ {\scriptscriptstyle\square} \ s_j \big>  \\  
  \end{array} \right \}
\end{displaymath}

The symbol ${\scriptscriptstyle\square}$ stands for any binary connective.
\end{definitionMinha}

\begin{displaymath}
p(s) =
\left\{ 
  \begin{array}{ccc}
  \Bar{p}(v) & \textrm{if} & \textrm{ ddddd } s =  v  \\
  p(s_i) \cdot p(s_j) & \textrm{if} & s = \big< s_i \ {\scriptscriptstyle\square} \ s_j \big>  \\  
  \end{array} \right \}
\end{displaymath}

The $\overline{p}$ function is applied initially to truth valuations by attributing a real numeric value to it.
The $p$ function encompasses all \textit{WFS} Definição \ref{def:SequenceBemFormada}.

Our computation revolves around the weight of a well-formed sequence, rather than that of a formula. Once extracted from a \textit{WFF} (Well-Formed Formula), the associated sequence no longer contributes to the respective formula's logical connectives. This contribution was already accounted for when the sequence was included in the justifying set. By categorizing elementary events as atomic \textit{WFS} and non-elementary events as elaborated \textit{WFS}, we make use of Definitions 3.8, 3.9, and 3.10 to establish a measure over logical sentences.

\begin{definitionMinha}\label{def:LogicalMeasure}
		[Logical Measure]: Let ${J}_{\alpha}$ be the set of well-formed sequences that justify $\alpha$, and  ${A}_\alpha$
		be the set of all well-formed sequences associated to $\alpha$. We call $Pl:{A}_\alpha \rightarrow [0 , 1]$,  
		the Logical Measure of $\alpha$ defined as

\begin{displaymath}
\mu(\alpha) = 		\Sigma_{\{ s \in {J}_{\alpha}\}}p(s)
\end{displaymath}
\end{definitionMinha}

For any $\alpha$, ${J}_{\alpha} \subseteq {A}_{\alpha}$ we have $\Sigma_{\{ s \in {A}_{\alpha}\}}p(s) = 1$, logo  $\Sigma_{\{ s \in {J}_{\alpha}\}}p(s) \leq 1$.
As $p(\cdot)$ is a positive map, $\Sigma_{\{ s \in {J}_{\alpha}\}}p(s) \geq 0$.
Example \ref{Ex:LogicalMeasure03} explores the defined measure:

\vspace{0.2cm}

\begin{aragao}\label{Ex:LogicalMeasure03}[Logical Measure 03] Let $\varphi$ be $(A \vee \neg B)$ and ${C} = \{ v_0, v_1, v_2, v_3 \}$. 
The definition of Logical Measure
relies on the definitions of justification and association, however,
as we saw in the Example (\ref{ex:ElementaryAleatoryEventOfaFormulaDISJUNCAO}),
$ {A_{(A \vee B)}} = \{ \big<v_i , v_j\big> \ | \ v_i(A) = 1 \textrm{ and } v_j(B) = 0
\}$
and 
$ {J_{(A \vee B)}} = \{
\big< v_2  ,   v_0 \big>\ ,
\big<
 v_2  ,   v_2 
\big>\ , \ 
\big< v_3  ,   v_0 \big>\ ,
\big<
 v_3   ,  v_2  \big>$\}.
Applying Definition \ref{x} we get the Logical Measure of $\varphi$ as
\textcolor{black}{
$
\mu(\varphi) = 		\Sigma_{\{ s \in {J}_{\varphi}\}}p(s) =
p(\big< v_1 \vee  v_2 \big>) +
p(\big< v_2 \vee  v_3 \big>) + 
p(\big< v_3 \vee  v_2 \big>\ ,
p(\big< v_3 \vee v_3  \big>)$
}.

\end{aragao}

\textcolor{black}{
During the justification phase, the connectives were factored in. However, during the process of Weight Allotment, this consideration is no longer necessary. We simply compute their values in accordance with the fundamental principle of counting.
}

\subsubsection{Conclusion}\label{sec:Conclusion}

We have formulated a formal system that combines the characteristics of both logic and probability theory. PML (Propositional Measure Logic) encapsulates the advantages of both realms, as it remains accurate and comprehensive in its adherence to the semantics of Classical Propositional Logic. Consequently, the deductive inference inherent in logic is preserved within PML, whose semantic framework is founded upon mathematical measures, leveraging the quantitative nuances of a metric such as Probability. PML inherits the reliability of sound logical deductive reasoning while also addressing context-sensitive beliefs through probability, offering a succinct encapsulation of

Before we explain the deductive system we need to present Probabistic Entailment.

\section{Probabistic Entailment}

\begin{definition}[Probabilistic Entailment]
We say that \( Q \) is a probabilistic consequence of \( P \) and \( (P \rightarrow Q) \), denoted by  
\[ \{P, \ (P \rightarrow Q)\} \models Q, \]  
when \( \mu(P) \) and \( \mu(P \rightarrow Q) \) exceed a threshold \( k \), we find that \( \mu(Q) \) also exceeds \( k \).  
\end{definition}

By manipulating the definition of the measure $\mu$, we find that 

\begin{itemize}
    \item $v_2 \geq k/2$ and $v_4 \geq k/2$
    \item Or at least one of them close to $k$ while the other does not tend to zero. 
\end{itemize}

Unlike Nilsson's logic, the entailment in PML is not represented by a convex hull, but rather by a surface. 

\section{Soudness}

Probabilistic semantics is a probabilistically consistent extension of classical propositional logic, since it preserves the positivity of logical truths.

\begin{theorem}[Probabilistic Soundness]
If $\models_\text{CL} \varphi$ ($\varphi$ is a classical tautology), then $P(\varphi) > 0$ in the proposed semantics.
\end{theorem}

\subsection{Adapted Definition}

Based on Nilsson and Halperin\footnote{
“A logic is probabilistically sound if the probability of the conclusion is at least as high as the joint probability of the premises when they exceed a certain threshold.”
(In this article, Nilsson discusses the generalization of logical implication to probabilistic scenarios, where correctness is defined in terms of preserving probabilistic bounds).  Nilsson, N. J. (1986). “Probabilistic Logic”, Artificial Intelligence, 28(1), pp. 71-87.
Page 75.
“A probabilistic inference system is sound if it does not allow conclusions to be derived with probability less than the joint probability of the premises when they reach a specified threshold.” Halpern, J. Y. (2003). “Reasoning About Uncertainty”, MIT Press. Chapter 7, Page 215.
}, we define:

\begin{definition}[Probabilistic Correctness][\cite{nilsson}]
$\mathcal{L}_P$ is \textbf{correct} if for every valid inference $\Gamma \vdash_P \varphi$, the conclusion $\varphi$ has probability not less than the premises when $\mu(\Gamma) \geq k$.
\end{definition}

\noindent Based on Fagin and Ognjanović\footnote{
 “A Logic for Reasoning About Probabilities”, Information and Computation, 87(1-2), pp. 78-128.

Page 83:
“A probabilistic logic is complete if every consequence that holds with probability above a threshold (given premises above that threshold) is probable in the system.” Fagin, R., Halpern, J. Y., \& Megiddo, N. (1990).
“Completeness in probabilistic systems requires that all semantically valid conclusions of high probability be syntactically derivable.” Ognjanović, Z., $\&$ Rašković, M. (2000). Page 792: paper “Some Probability Logics with New Types of Probability Operators”, Journal of Logic and Computation, 10(6), pp. 787-808.
}, we define:

\begin{theorem}[Classical Connection]
Let $\mathcal{L}_C$ be classical propositional logic. For $k = 1$:
\begin{align*}
\text{if } \mathcal{L}_P \text{is:} &
\begin{cases}
\text{correct} & \Rightarrow \mathcal{L}_P \text{ generalizes } \mathcal{L}_C \\
\text{complete} & \Rightarrow \mathcal{L}_C \text{ is a particular case of } \mathcal{L}_P
\end{cases}
\end{align*}
\end{theorem}

\subsection{Soudness}

Every true formula has positive measure.

\begin {proof}(Soudness) 
Induction testing in the \textit{wff} structure:
\end {proof}

If a formula is true, then it is justified i.e. there is a $ s $ sequence that justifies it.

\noindent [Base Case] \ 
$A$ is true $\Rightarrow \ \  \mu (A)	> 0$. 

\noindent By definition, 
$A$ be true implies there exist $v^*$ which $v^*(A) = 1$ and $p(v^*) > 0$.  

Just pick up\footnote{Remember if there only one propositional letter $A$, we have just two possible truth valuation, $v_1$ and $v_2$. Here, $v_2$ is the case.}  $\big< v^* \big>$ to be $s$. 
As $v^*(A) = 1$, $\big< v^* \big> \in {J}_A$ and by Definition \ref{def:justificacaoBis}, we have
              $\big< s \big> \ |\!\!\! \approx \  (A)$, then $\mu(A)$ is greater than zero.

\noindent [Induction Hypothesis]
\noindent $\alpha$ is true $\Rightarrow \ \  \mu (\alpha) > 0$

\noindent [Inductive Step: Negation]
\noindent Let $(\neg \alpha)$ true. 
by Definition \ref{def:justificacaoBis}, $\Rightarrow$ exist $v^*$, $v^*(\alpha) = 0$,
Let $\big< v^* \big>$ be $s$. 
               By lemma 01, we have,
              $\big< s \big> \ |\!\!\! \not \approx \  \alpha$

by Definition \ref{def:justificacaoBis}
$\big< s \big> \ |\!\!\! \approx \  (\neg \alpha)$
then $\mu(\neg \alpha) > 0$.
 $\square$

\noindent [Inductive Step conjunction]			
\noindent   Let $\bullet$ $A \wedge B$ be TRUE, then exists $v^*(\alpha \wedge \beta) = 1$  \

            thus  $v^{A \wedge B}(A) = 1$ and
$v^{A \wedge B}(B) = 1$.

$\bullet$ Therefore $\big< v^{A \wedge B}\big> \in  {J}_A$ and 
$\big< v^{A \wedge B}\big> \in  {J}_B$.

Thus, $s = \big< v^{A \wedge B},  v^{A \wedge B}\big>$ belongs to 
${J}_{(A \wedge B)}$, (By Definition)

so, we conclude that $\mu(A \wedge B) > 0$.
  $\square$

\subsection{About Completness}

Since the measure captures belief i.e. degrees of truth, 
it is not expected to correspond to the absolute truth of classical logic. Therefore, not every probable formula will be classically true.

Completeness between classical semantics and probabilistic semantics is not desirable.
For example, an expert committee where some advocate a hypothesis and others advocate against it. This hypothesis has a degree of belief other than zero, but it is not absolute truth (classical) because it does not represent unity.

\section{Related Work}
\label{sec:related-work}

Works involving Probability and Logic can be classified into 3 branches

\textbf{Classical Probabilistic Logic} - 
The early atempts, \cite{gaifman1964concerning} and \cite{scott1964probability}, to assign probability measures to logical formulas were limited to \emph{truth-functional} probabilities. 
Later, \cite{halpern1990axiomatization}
extended this idea by using possible-worlds semantics, increasing the capacity for probabilistic interpretation.

\textbf{Probabilistic Extensions of Logical Frameworks} - 

There were also efforts linking Probability with First Order Logic, like 
   \emph{Probabilistic Propositional Logic} (PPL) by \cite{nilsson1986probabilistic}, which computes probabilities via linear constraints on possible worlds.
       \emph{Markov Logic Networks} \cite{richardson2006markov}, combining first-order logic with probabilistic graphical models to handle relational uncertainty.

\textbf{Probability and Non-Classical Logics} - 
Works about Probability and non-classical logics:

\cite{weatherson2003many} explored probabilistic interpretations of supervaluationism.
    \cite{bradley2016imprecise} developed imprecise probabilities for multi-valued logics.

\section{Future Work}
\label{sec:future-work}

Our probabilistic semantics for propositional logic opens several research directions:

\subsection{Generalization to First-Order Logic}
Extending the current framework to quantify over domains while preserving probabilistic soundness:
\begin{itemize}
    \item Define probability measures over predicate valuations.
    \item Investigate if $\models_{\text{FO}} \varphi$ implies $P(\varphi) > 0$ for first-order tautologies.
\end{itemize}

\subsection{Algorithmic Applications}
\begin{itemize}
    \item \textbf{Probabilistic SAT Solving}: Develop SAT solvers that compute $P(\varphi)$ for CNF formulas.
    \item \textbf{Learning}: Infer probabilistic logical rules from data (e.g., via Markov Logic Networks \cite{richardson2006markov}).
\end{itemize}

\subsection{Connections to Non-Classical Logics}
\begin{itemize}
    \item \textbf{Intuitionistic Logic}: Characterize probabilistic semantics for constructive proofs.
    \item \textbf{Quantum Probability}: Explore mappings between quantum states and formula probabilities.
\end{itemize}

\subsection{Axiomatization}
\begin{itemize}
    \item Derive a complete axiomatic system for reasoning about $P(\varphi)$ directly.
    \item Compare with Hájek's axiomatization for probabilistic modal logic \cite{hajek}.
\end{itemize}

\begin{enumerate}
    \item Define the proof system $\vdash_P$,
    \item Check whether $\mathcal{L}_C$ is recovered for $k=1$,
    \item Investigate properties for $0 < k < 1$.
\end{enumerate}

\section{Conclusão}

We have proved that probabilistic semantics is classically sound: every propositional tautology has positive probability, preserving logical truths.

\bibliographystyle{plain}

\end{document}